\documentclass[twocolumn,twoside,showpacs,superscriptaddress,nofootinbib,pdftex]{revtex4}

\usepackage[dvips]{graphicx}
\usepackage[dvips]{color}
\usepackage{epsfig}
\usepackage{amssymb,amsmath}
\usepackage{lscape}
\usepackage{color}

\RequirePackage{xspace}

\newcommand{\dsdpi}{\ensuremath{D^{*+}\to D^0\pi^+}\xspace}

\newcommand{\cconj}{\ensuremath{\mathcal{C}}\xspace}
\newcommand{\cpconj}{\ensuremath{\mathcal{CP}}\xspace}

\newcommand{\kspp}{\ensuremath{K^0_S\pi^+\pi^-}\xspace}
\newcommand{\dkpp}{\ensuremath{D\to K^0_S\pi^+\pi^-}\xspace}
\newcommand{\dnkpp}{\ensuremath{D^0\to K^0_S\pi^+\pi^-}\xspace}
\newcommand{\dbkpp}{\ensuremath{\overline{D}{}^0\to K^0_S\pi^+\pi^-}\xspace}

\newcommand{\bdk}{\ensuremath{B^+\to D K^+}\xspace}

\newcommand{\dn}{\ensuremath{D^0}\xspace}
\newcommand{\dnbar}{\ensuremath{\overline{D}{}^0}\xspace}

\newcommand{\ab}{\ensuremath{A_B}\xspace}

\newcommand{\pd}{\ensuremath{P}\xspace}
\newcommand{\pdbar}{\ensuremath{\overline{P}}\xspace}

\newcommand{\ad}{\ensuremath{A}\xspace}
\newcommand{\adbar}{\ensuremath{\overline{A}}\xspace}
\newcommand{\dvar}{\ensuremath{(m^2_{+}, m^2_{-})}\xspace}

\newcommand{\aad}{\ensuremath{|A|}\xspace}
\newcommand{\aadbar}{\ensuremath{|\overline{A}|}\xspace}

\begin{document}

\title{Charm mixing in the model-independent analysis of 
       correlated $\mathbf{\dn\dnbar}$ decays}

\author{Alex Bondar}
\affiliation{Budker Institute of Nuclear Physics, 11 Lavrentieva, Novosibirsk, 630090, Russia}
\affiliation{Novosibirsk State University, 2 Pirogova, 630090, Russia}
\author{Anton Poluektov}
\affiliation{Budker Institute of Nuclear Physics, 11 Lavrentieva, Novosibirsk, 630090, Russia}
\affiliation{Department of Physics, University of Warwick, Coventry CV4 7AL, United Kingdom}
\author{Vitaly Vorobiev}
\affiliation{Budker Institute of Nuclear Physics, 11 Lavrentieva, Novosibirsk, 630090, Russia}
\affiliation{Novosibirsk State University, 2 Pirogova, 630090, Russia}

\date{\today}

\begin{abstract}
We investigate the impact of charm mixing on the 
model-independent $\gamma$ measurement using Dalitz plot 
analysis of the three-body $D$ decay from \bdk process,
and show that ignoring the 
mixing at all stages of the analysis is safe up to a sub-degree
level of precision. We also find that in the coherent production of the 
$\dn \overline{D}{}^{*0}$ system in $e^+e^-$ collisions, the effect of 
charm mixing is enhanced, and propose a model-independent method to 
measure charm mixing parameters in time-integrated Dalitz plot analysis at 
charm factories. 
\end{abstract}

\pacs{13.25.Hw, 13.25.Ft, 12.15.Hh, 11.30.Er}

\maketitle

\section{Introduction}

Dalitz plot analysis of three-body decays of neutral $D$ mesons is a useful tool 
in various measurements where coherent admixtures of \dn and \dnbar are 
observed. This technique was initially proposed for the measurement 
of the unitarity triangle angle $\gamma$ in \bdk decays \cite{giri,bondar}. 
Later it was applied to the measurement of charm 
mixing~\cite{belle_dmix_dalitz, babar_dmix_dalitz} and to the 
resolution of the quadratic ambiguity in the measurement 
of the angle $\beta$ using a time-dependent analysis of the decay 
$B^0\to D\pi^0$~\cite{Dpi0,belle_dpi0}. 
Most of these measurements are based on the \dkpp decay which 
offers the best precision among three-body $D^0$ decays. 

The technique is model-dependent --- it depends on the complex amplitude 
of the \dn decay which is obtained from the \dsdpi sample using model assumptions. 
The result of the measurement contains therefore model uncertainties. 
In the case of $\gamma$ measurement, this uncertainty ($\sim 10^{\circ}$) 
is already comparable to the statistical 
accuracy~\cite{belle_phi3, babar_gamma}. 

However, a modification of the analysis is possible that allows to 
perform a completely model-independent measurement~\cite{giri}. 
It requires the 
phase space of the three-body $D$ decay to be divided into bins. 
Information about the complex phase in each bin can be extracted from 
the quantum-correlated $D^0$ decays from $\psi(3770)\to D\bar{D}$ process. 
The measurement of the strong phase in bins of the \dkpp phase space
was recently performed by the CLEO collaboration~\cite{cleo_dkpp_corr}. 
This measurement should 
allow to reduce the error of $\gamma$ related to the uncertainty in the 
\dkpp amplitude to $1-2^{\circ}$. 

Recently, charm mixing was observed by the Belle and BaBar 
experiments~\cite{belle_mixing, babar_mixing}.
With degree-level precision, the effect of charm mixing can become 
significant in the measurement of $\gamma$. It was shown that mixing 
contributes only of second order in the $x$ and $y$ parameters 
to the ADS, GLW and model-dependent Dalitz plot analysis methods, and thus can 
be safely neglected~\cite{gamma_mix}. But the effect of mixing on the binned analysis 
with the phase terms extracted from quantum-correlated $D^0$ decays 
is of separate interest. 

In this paper, we investigate the impact of the charm mixing on the 
model-independent $\gamma$ measurement, and show that ignoring the 
mixing at all stages of the analysis is safe up to a sub-degree
level of precision. We also find that in the coherent production of 
$\dn \overline{D}{}^{*0}$ system in $e^+e^-$ collisions, the effect of 
charm mixing is enhanced compared to the case of $\dn\dnbar$ production, 
and propose a model-independent method to measure charm mixing 
parameters in time-integrated Dalitz analysis at 
charm factories. The method is sensitive to both mixing parameters, 
$x$ and $y$, as well as to \cpconj violation parameters $r_{\cpconj}$  and 
$\alpha_{\cpconj}$. 
The sensitivity of the proposed method can be improved by adding 
doubly Cabibbo-suppressed three-body modes such as $D^0\to K^+\pi^-\pi^0$. 
We estimate the sensitivity of the proposed method using Monte-Carlo (MC)
simulation. 

\section{Model-independent binned analysis of three-body $\mathbf{\dn}$ decays}

To introduce the notation we briefly recap the technique of 
model-independent binned Dalitz plot analysis of \bdk, \dkpp decays 
used to extract the angle $\gamma$. As usually presented, this does not 
take charm mixing effects into account.

The amplitude of the \bdk, \dkpp decay can be written as
\begin{equation}
  \ab=\adbar+ r_Be^{i(\delta_B+\gamma)}\ad\,, 
\end{equation}
where $\adbar=\adbar(m^2_{K_S\pi^+}, m^2_{K_S\pi^-})\equiv\adbar\dvar$ 
is the amplitude of the \dbkpp decay,  $\ad=\ad\dvar$ 
is the amplitude of the \dnkpp decay ($\ad\dvar = \adbar(m^2_{-},m^2_{+})$ 
in the case of \cpconj conservation in $D$ decay), 
$r_B$ is the ratio of the absolute values of the interfering 
$B^+\to \overline{D}{}^0K^+$ and $B^+\to D^0K^+$ amplitudes, 
and $\delta_B$ is the strong phase difference between these amplitudes. 
The density of the $D$ decay Dalitz plot from \bdk\ decay is given by the 
absolute value squared of the amplitude 
\begin{equation}
 \begin{split}
  P_{B}=|\ab|^2 = & |\adbar+ r_Be^{i(\delta_B+\gamma)}\ad|^2 =   
  \label{p_b} \\
  & \pdbar +r_B^2\pd + 2\sqrt{\pd\pdbar}(x_B C+y_B S)\,, 
 \end{split}
\end{equation}
where 
\begin{equation}
  x_B = r_B\cos(\delta_B+\gamma)\,; \;\;\;
  y_B = r_B\sin(\delta_B+\gamma)\,.
\end{equation}
The functions $C=C\dvar$ and $S=S\dvar$ are the cosine and sine of the strong 
phase difference $\delta_D=\arg\adbar-\arg\ad$ between the \dbkpp and \dnkpp 
amplitudes\footnote{
  This paper follows the convention for strong phases in $D$ decay
  amplitudes introduced in Ref.~\cite{modind2008}. 
}: 
\begin{equation}
  C=\cos\delta_D(m^2_+,m^2_-)\,; \;\;\;
  S=\sin\delta_D(m^2_+,m^2_-)\,. 
\end{equation}
The equations for the charge-conjugate mode $B^-\to D K^-$ are obtained 
with the substitution $\gamma \longrightarrow -\gamma$. Using both 
$B$ charges, one can obtain $\gamma$ and $\delta_B$ separately. 

In the binned model-independent approach, the Dalitz plot is divided into 
$2\mathcal{N}$ bins symmetrically to the exchange $m^2_-\leftrightarrow m^2_+$. 
The expected number of events in the bin ``$i$'' of the Dalitz plot of
$D$ from \bdk\ is
\begin{equation}
  N_i = 
  h_{B} \left[
    K_i + r_B^2K_{-i} + 2\sqrt{K_iK_{-i}}(x_B C_i+y_B S_i)
  \right] \,, 
  \label{n_b}
\end{equation}
where $K_i$ is the number of events in the corresponding bin of the Dalitz plot 
of the $D$ meson in a flavor eigenstate (obtained using
$D^{*\pm}\to D\pi^\pm$ samples) and $h_{B}$ is a normalization constant. 
The bin index ``$i$'' ranges from $-\mathcal{N}$ to $\mathcal{N}$ (excluding 0); 
the exchange $m^2_+ \leftrightarrow m^2_-$ corresponds to the exchange 
$i\leftrightarrow -i$. 
The terms $C_i$ and $S_i$ include information about 
the cosine and sine of the phase difference averaged over the bin region:
\begin{equation}
  C_i=\frac{\int\limits_{\mathcal{D}_i}
            \aad\aadbar
            \cos\delta_D\,d\mathcal{D}
            }{\sqrt{
            \int\limits_{\mathcal{D}_i}\aad^2 d\mathcal{D}
            \int\limits_{\mathcal{D}_i}\aadbar^2 d\mathcal{D}
            }}\,. 
  \label{cs}
\end{equation}
Here $\mathcal{D}$ represents the Dalitz plot phase space and 
$\mathcal{D}_i$ is the bin region over which the integration is performed. 
The terms $S_i$ are defined similarly with cosine substituted by sine. 

The expected number of events in each Dalitz plot 
bin (\ref{n_b}) is trivially obtained 
from the probability density (\ref{p_b}) by integrating 
over the bin area, which leads to the substitutions 
$\pd\to K_i$, $\pdbar\to K_{-i}$, $C,S\to C_i, S_i$. 
In what follows, we only quote the number of events to save space. 
Normalization constants (such as $h_B$ in (\ref{n_b})) are also omitted.

The symmetry under $\pi^+ \leftrightarrow \pi^-$ requires $C_i = C_{-i}$ and
$S_i = - S_{-i}$. The values of $C_i$ and $S_i$ terms can be provided by
charm-factory experiments operated at the threshold of 
$D\overline{D}$ pair production~\cite{cleo_dkpp_corr}. 
The wave function of the two mesons is antisymmetric, thus 
the four-dimensional density of two correlated \dkpp Dalitz plots is
\begin{equation}
\begin{split}
  |A_{\rm corr}&(m_+^2,m_-^2,m'^2_+,m'^2_-)|^2=|\ad_1\adbar_2-\adbar_1\ad_2|^2=\\
     &\pd_1\pdbar_2 + \pdbar_1 \pd_2 -
     2\sqrt{\pd_1\pdbar_2\pdbar_1\pd_2}(C_1 C_2+S_1 S_2)\,, 
  \label{p_corr}
\end{split}
\end{equation}
where the indices ``1'' and ``2'' correspond to the two decaying $D$ mesons. 
In the case of a binned analysis, the number of events in the region of the
$(K^0_S\pi^+\pi^-)^2$ phase space described by the indices ``$i$'' and ``$j$'' is 
\begin{equation}
\begin{split}
  M_{ij} = &K_i K_{-j} + K_{-i} K_j  \\
    -&2\sqrt{K_iK_{-i}K_jK_{-j}}(C_i C_j + S_i S_j). 
\end{split}
\end{equation}
Once the values of the terms $C_i$ and $S_i$ are known from 
charm-factory data, the system of equations (\ref{n_b})
contains only three free parameters ($x_B$, $y_B$, and $h_B$)
for each $B$ charge, and can be solved using maximum likelihood 
method to extract the value of $\gamma$. 

Note that technically the system (\ref{n_b}) can be solved without 
external constraints on $C_i$ and $S_i$ for $\mathcal{N} \ge 2$.
However, due to the small value of $r_B$, there is very little
sensitivity to the $C_i$ and $S_i$ parameters in \bdk decays, which 
results in a reduction in the precision on $\gamma$ that can be 
obtained~\cite{modind2006}.

\section{Contribution of charm mixing to model-independent $\mathbf{\gamma}$ measurement.}\label{gamma_mixing}

In the case of \cpconj conservation, the mass eigenstates of the neutral $D$
system are given by 
\begin{equation}
  D_{1,2}=\frac{1}{\sqrt{2}}(\dn\pm\dnbar).
  \label{d_cpcons}
\end{equation}

Charm mixing is described by two parameters, $x_D$ and $y_D$, 
which are defined as
\begin{equation}
  x_D=\frac{m_2-m_1}{\Gamma},\quad y_D=\frac{\Gamma_2-\Gamma_1}{2\Gamma}, 
\end{equation}
where $m_{1,2}$ and $\Gamma_{1,2}$ are the mass and decay widths of the 
mass eigenstates. 
We use notations $x_D$ and $y_D$ instead of the more common $x$ and $y$ 
in order not to confuse 
them with the \cpconj-violating parameters $x_B$ and $y_B$ introduced before. 
The current world average values are: 
$x_D=(0.59\pm 0.20)\%$, 
$y_D=(0.80\pm 0.13)\%$~\cite{hfag_mixing}.

\cpconj violation modifies the expression~(\ref{d_cpcons}) to 
\begin{equation}
  D_{1,2}=p\dn\pm q\dnbar, 
  \label{d_cpv}
\end{equation}
where $p$ and $q$ satisfy $|p|^2+|q|^2=1$. 
\cpconj-violating mixing is thus described by two additional parameters
$r_{\cpconj}$ and $\alpha_{\cpconj}$: 
\begin{equation}
  r_{\cpconj}e^{i\alpha_{\cpconj}}=q/p. 
\end{equation}

Below we present all the quantities that enter the model-independent 
analysis including the contribution of the \cpconj-conserving charm mixing 
(the corresponding quantities are denoted with the prime mark). 
The full formalism including \cpconj violation in mixing is given in 
Appendix~\ref{full_formalism}.

The number of events in the each bin of the flavor-tagged \dnkpp Dalitz plot 
after integration over decay time is 
\begin{equation}
  K'_i = K_i + \sqrt{K_i K_{-i}}(y_D C_i+x_D S_i) +
  		 O\left (x^2_D,y^2_D\right ).
  \label{k_mix}
\end{equation}

Similarly, one can obtain the number of events for the \dkpp decay from \bdk: 
\begin{equation}
  \begin{split}
    N'_i = & N_i+\sqrt{K_iK_{-i}}\left (y_DC_i+x_DS_i\right )\\
           +& r_B^2\sqrt{K_iK_{-i}}\left (y_DC_i-x_DS_i\right )\\
           +& K_i\left (x_By_D-y_Bx_D\right )\\
           +& K_{-i}\left (x_By_D+y_Bx_D\right )+O\left (x^2_D,y^2_D\right ). 
  \label{n_mix}
  \end{split}
\end{equation}

The amplitude for the correlated $D^0\overline{D}{}^0$
decay, assuming that the particle denoted with the index ``1'' decayed 
first, equals 
\begin{equation}
  A_{\rm corr}(t,m_+^2,m_-^2,m'^2_+,m'^2_-)=\ad_1(0)\adbar_2(t)-\adbar_1(0)\ad_2(t). 
  \label{a_corr}
\end{equation}
Before the particle ``1'' decays, the amplitude stays antisymmetric and 
the mixing does not affect it. We can therefore assume that the particle ``1''
decays at the time $t=0$ and count time from the moment of its decay. 

After the integration over time and taking the absolute value squared of the 
amplitude~(\ref{a_corr}), we obtain the number of events $M'_{ij}$ in the bin ``$ij$''
of the phase space for the pair of $D$ mesons (still assuming that the 
particle ``1'' denoted here with the index ``$i$'' decayed first):
\begin{equation}
\begin{split}
  M'_{ij} = & K_i K_{-j}+K_{-i}K_j \\
  -& 2\sqrt{K_iK_{-i}K_j K_{-j}}(C_iC_j + S_iS_j)\\
  -& K_j\sqrt{K_iK_{-i}}(y_DC_i-x_DS_i)\\
  -& K_{-j}\sqrt{K_iK_{-i}}(y_DC_i+x_DS_i)\\
  +& K_i\sqrt{K_j K_{-j}}(y_DC_j-x_DS_j)\\
  +& K_{-i}\sqrt{K_j K_{-j}}(y_DC_j+x_DS_j)\\
  +& O(x_D^2, y_D^2).
\end{split}
\end{equation}
However, measurement of the decay time in an experiment with 
symmetric beams is difficult. Therefore one has to average over 
the decay order, which leads to the cancellation of all terms 
linear in mixing parameters: 
\begin{equation}
  M'_{ij} = M_{ij} + O(x_D^2, y_D^2). 
  \label{m_mix}
\end{equation}

The real analysis performed at CLEO uses the values $M'_{ij}$ (which, as
we have seen, are unaffected by mixing at first order), and values of $K_i$ 
obtained from correlated $D\overline{D}$ decays where one of the $D$ mesons serves as a flavor 
tag. The $K_i$ values extracted this way are also unaffected by mixing. 
Therefore, the values of $C_i$ and $S_i$ extracted in this analysis 
contain no linear mixing contribution. 

As far as processes observed at $B$ factories are concerned, both the 
\dkpp decay from \bdk and the flavor-tagged \dkpp decay contain mixing 
contributions, and therefore the observable numbers of events in the 
Dalitz plot bins are $N'_i$ and $K'_i$, respectively. Clearly, if one 
uses the values $K_i$, obtained from a charm factory, in the fit to 
obtain $\gamma$ from $N'_i$ described by Eq.~\ref{n_mix}, the resulting
value contains contribution in first order in $x_D$, $y_D$. 
If the values $K'_i$ are used, Eq.~\ref{n_mix} can be rewritten as
\begin{equation}
  N'_i = 
    K'_i + r_B^2K'_{-i} + 2\sqrt{K'_iK'_{-i}}(x_B C'_i+y_B S'_i)+O(x_D^2,y_D^2)
  \,, 
  \label{n_mix_k_mix}
\end{equation}
{\it i. e. } it has the same form as Eq.~\ref{n_b}, without mixing 
contributions up to second-order in $x_D,y_D$, but the 
phase terms $C'_i, S'_i$ contain first-order mixing corrections:
\begin{equation}
\begin{split}
  C'_i = C_i & + \frac{K'_i+K'_{-i}}{\sqrt{K'_i K'_{-i}}} (1-C^2_i)y_D
               + \frac{K'_i-K'_{-i}}{\sqrt{K'_i K'_{-i}}} C_iS_ix_D, \\
  S'_i = S_i & - \frac{K'_i-K'_{-i}}{\sqrt{K'_i K'_{-i}}} (1-S^2_i)x_D
               - \frac{K'_i+K'_{-i}}{\sqrt{K'_i K'_{-i}}} C_iS_iy_D.
\end{split}
\end{equation}

Thus, if the terms $C_i,S_i$ are left as free parameters in the fit to $B$ decay 
data, the mixing correction is only of second order (the effective parameters 
$C'_i$, $S'_i$ are measured in this case \cite{gamma_mix}), 
but if these terms are obtained from 
correlated $D$ decays, first order mixing corrections to $\gamma$ appear. 
However, these corrections are additionally suppressed by factor $r_B\sim 0.1$, 
and the residual contribution of charm mixing to $x_B$ and $y_B$ 
is at a percent level.

A quantitative estimate of the effect was performed using the procedure 
described in Appendix~\ref{num_calc}. 
Three analysis strategies have been considered: 
\begin{enumerate}
  \item Using $K_i$ (unaffected by mixing) from the coherent $D\overline{D}$ production,
  \item Using $K'_i$ (Eq.~\ref{k_mix}) measured in \dsdpi, \dkpp decays,
  \item Using $K_i$ and applying a linear correction for the mixing 
        contribution according to Eq.~\ref{n_mix}
        (assuming that $x_D$, $y_D$ are known). 
\end{enumerate}
The effect of mixing on the fitted value of $\gamma$ depends on the value
$\alpha_D=\arctan(y_D/x_D)$, the ratio $\sqrt{x_D^2+y_D^2}/r_B$, 
and the $\delta_B$ and $\gamma$ values. 
In our study, we use $\sqrt{x_D^2+y_D^2}/r_B=0.1$ (all biases are 
proportional to this quantity) and 
scan over the other parameters. The results are shown in Table~\ref{tab:Gamma_bias}. 
Clearly, if the $K'_i$ values from \dsdpi, \dkpp decays are used in 
the $\gamma$ fit, the mixing contribution can be neglected.

\begin{table}
\begin{center}
\caption{Estimates of the charm mixing effect on the $\gamma$ value measured using the
	 model-independent time-integrated Dalitz plot analysis method, for three 
         different analysis strategies. 
         The maximum $\gamma$ biases for varying $\alpha_{D}=\arctan(y_D/x_D)$, 
         $\delta_B$ and $\gamma$, as well as the values of these 
         parameters at which the maximum is reached, are shown. 
         The estimation assumes $\sqrt{x_D^2+y_D^2}/r_B=0.1$.}
\label{tab:Gamma_bias}
\begin{tabular}{|l|c|c|c|c|}
\hline
Strategy               &$\Delta\gamma_{\rm max}$&$\alpha_{\rm max}$&$\delta_{B,
\rm max}$&$\gamma_{B, \rm max}$\\ 
\hline 
1. Using $K_i$         &$2.9^{\circ}$  &$184^{\circ}$ &$85^{\circ}$ &$87^{\circ}$\\ 
2. Using $K_i^{\prime}$&$-0.2^{\circ}$ &$97^{\circ}$  &$2^{\circ}$  &$90^{\circ}$\\ 
3. Linear correction   &$0.07^{\circ}$ &$324^{\circ}$ &$72^{\circ}$ &$73^{\circ}$\\ 
\hline
\end{tabular}
\end{center}
\end{table}

For clarity, we would like to compare our result with the conclusions 
of previous papers that have addressed the impact of charm mixing on 
$\gamma$ measurements~\cite{gamma_mix, silva}. Ref.~\cite{silva} considers
the case when the $D$ amplitude does not contain the mixing contribution (or 
is corrected for it), but the $B$ decay data are uncorrected for mixing. 
This corresponds to the analysis strategy 1 in our MC study.
The systematic bias to $\gamma$ in that case is 
linear in $x_D$ and $y_D$ and can be numerically large. The treatment
in~\cite{gamma_mix} corresponds to the case when the mixing is neglected 
in both the flavor-tagged $D$ and $B$ data; the systematic bias in 
$\gamma$ is second order in $x_D$ and $y_D$ in that case. In the 
context of the model-independent binned Dalitz plot analysis, the conclusions 
of~\cite{gamma_mix} can only be applied if the phase terms $C_i$ and $S_i$
are left as free parameters in the fit to $B$ data. The analysis procedure 
considered here is an intermediate case: part of information about 
the \dkpp amplitude (namely, $K_i$ terms) is extracted from the flavor-tagged
data uncorrected for mixing, while the $C_i$ and $S_i$ terms from the 
quantum-correlated $D\overline{D}$ decays contain no mixing contribution. This 
results in a bias linear in $x_D, y_D$ (but numerically small due to 
additional $r_B$ suppression). 

\section{Time-dependent charm mixing measurement}

\label{mixing_tdep}

The charm sector is the only place where contributions to \cpconj 
violation from down-type quarks in the mixing diagram can be explored. 
While values of charm mixing parameters are not easy to predict 
in the SM, \cpconj violation in mixing is expected to be very small. 
However, there is a range of SM extensions which allow sizable \cpconj violation 
effects~\cite{sm_extensions_grossman,sm_extensions_golowich}. 
Because of that, precise 
measurements of mixing as well as \cpconj violation parameters  
are essential.

The most accurate measurements of the charm mixing parameters to date 
have been performed by $B$-factories using time-dependent 
methods~\cite{belle_mixing, babar_mixing}. For example, by observing the 
wrong-sign decay $D^0\to K^+\pi^-$~\cite{babar_mixing}, BaBar determines 
$R_D$, the ratio of doubly Cabibbo-suppressed (DCS) to Cabibbo-favored (CF) 
decay rates, 
and the mixing parameters $x_D^{\prime\,2}$ and $y_D^{\prime}$, 
where $x_D^{\prime}=x_D\cos{\delta_{K\pi}}+y_D\sin{\delta_{K\pi}}$, 
$y_D^{\prime}=-x_D\sin{\delta_{K\pi}}+y_D\cos{\delta_{K\pi}}$ and 
$\delta_{K\pi}$ is the strong phase between the DCS and CF amplitudes. 
Since only $y_D^{\prime}$ contributes linearly to the time dependence 
of the decay rate, 
and the strong phase $\delta_{K\pi}$ is close to zero~\cite{cleo_dcs}, 
this kind of measurement is practically insensitive to the parameter $x_D$.

The Belle collaboration has performed a measurement of mixing parameters using 
time-dependent Dalitz plot analysis of the $D^0\to K_S^0\pi^+\pi^-$ 
decay~\cite{belle_dmix_dalitz}. A similar analysis has been performed by 
the BaBar collaboration using in addition $D^0\to K_S^0K^+K^-$~\cite{babar_dmix_dalitz}.
Here the Dalitz plot distribution depends linearly on both mixing parameters.
However, model assumptions in the decay amplitude 
description are necessary in this analysis, which unavoidably 
results in significant model uncertainties.

The model-independent binned Dalitz plot analysis may be extended to time-dependent 
measurements of charm mixing parameters and \cpconj violation in charm mixing. 
The time-dependent number of events in the bin ``$i$'' of $D^0\to K_S^0\pi^+\pi^-$ 
decay Dalitz plot is
\begin{equation}
\begin{split}
 K^{\prime}_i\left (t\right )=h_D e^{-\Gamma t}\left [ \right.
   & K_i+\sqrt{K_iK_{-i}}\left (y_DC_i+x_DS_i\right )\Gamma t+\\
   & \left. O\left (\left (x_D+y_D\right )^2\left (\Gamma t\right )^2\right )\right ].
\end{split}
\end{equation}

Using parameters $C_i$ and $S_i$ determined 
from independent measurements at a charm factory, we can eliminate completely
the model uncertainty in $x_D$ and $y_D$ extraction. 

\begin{table}
\begin{center}
\caption{Statistical sensitivity to the mixing and \cpconj violation
                 parameters for the time-dependent Dalitz plot analysis. 
                 Two strategies are considered: (i) $K_i$ fixed from charm
		 factory data and (ii) $K_i$ taken as free parameters.}
\label{tab:TimeDependentMeasurement}
\begin{tabular}{|l|c|c|}\hline
Parameter & \multicolumn{2}{|c|}{Precision} \\ \cline{2-3}

&$K_i$ fixed & $K_i$ floated\\ \hline
$x_D$ ($10^{-4}$)  			& $17$ & $22$ \\
$y_D$ ($10^{-4}$)  			& $13$ & $16$ \\
$r_{\cpconj}$ ($10^{-2}$)     		& $9$ & $9$ \\
$\alpha_{\cpconj}$ (${}^{\circ}$) 	& $5$ & $5$ \\ 
\hline
\end{tabular}
\end{center}
\end{table}

We estimate the statistical sensitivity of the time-dependent fit to the
mixing and the \cpconj violation parameters. Note that the values of 
$K_i$ can be measured independently with high precision at charm factory
experiments.
However, in that case one has to deal with systematic uncertainties due
to detector efficiency and other effects. Depending on the magnitude 
of these uncertainties, one can consider a fit with $K_i$
as free parameters, where the statistical uncertainty increases, but the 
systematic effects are minimized.

Table \ref{tab:TimeDependentMeasurement} shows the results for both strategies. 
The toy MC simulation uses $10^6$ flavor-tagged $D$ mesons from 
$D^*$ decays corresponding to the samples available at
the $B$-factories; the fit with fixed $K_i$ uses in addition 
$2\cdot 10^6$ flavor-tagged $D$ decays in $\cconj=-1$ state 
that can be obtained at a future charm factory experiment.
The simulation uses the amplitude of \dkpp decay determined by Belle
\cite{belle_phi3}, and the binning of the Dalitz plot with 8 bins defined by the 
uniform division of the strong phase difference $\delta_D$ \cite{modind2008}. 
The values of the phase terms $C_i, S_i$ are calculated from the \dkpp 
amplitude; the contribution of their statistical precision in the final 
result is negligible with the current experimental data.
Our estimates do not account for uncertainties in the time measurement and
background effects.

\section{Time-integrated charm mixing measurement}

\label{mixing_tint}

The cancellation of terms linear in $x_D$ and $y_D$ for the  
correlated decays occurs only in the case of the antisymmetric wave function
of two $D$ mesons produced with the charge conjugation
quantum number $\cconj=-1$. 
It is possible, however, to produce a pair of $D$ mesons
with both quantum numbers $\cconj=\pm 1$ in the process 
$e^+e^-\to\psi(4040)\to D^0 \overline{D}{}^{*0}$~\cite{asner_sun}. 
\footnote{The optimal energy for this kind of experiment is 
$4.01$ GeV, for which the $D\overline{D}{}^{*}$ cross section is maximized,
and $D^*\overline{D}{}^*$ production is below threshold.}
The $e^+e^-\to c\bar{c}$ cross-section at the $\psi(4040)$ mass is 
dominated by this process and is comparable to that at 
$\psi(3770)$~\cite{cleo_cs}.
Depending on the $D^{*0}$ final state ($D^0\pi^0$ or $D^0\gamma$), 
the $D\overline{D}$ pair is either $\cconj=-1$ or 
$\cconj=+1$, respectively. In the case of $\cconj=+1$, the amplitude 
of the decay is
\begin{equation}
  A_{\rm corr}=\ad_1\adbar_2+\adbar_1\ad_2. 
\end{equation}

Unlike Eq.~\ref{m_mix}, now we allow two $D$ mesons to decay into 
different final states, therefore we denote the numbers of events 
in bins of flavor-specific decays as $k_i$ and $K_i$, and the 
corresponding phase terms as $c_i, s_i$ and $C_i, S_i$. 
These can be either two-body (in that case only one bin makes sense, 
and the index ``$i$'' only takes the value $\pm 1$) 
or multibody final states. For non-self-conjugate final states, 
the index takes positive values for 
$D^0$ decay and negative values for $\overline{D}{}^0$ decay. 
The terms related to the strong phase difference
are defined similarly to $C_i$ and $S_i$ in Eq.~\ref{cs}. 
The number of events in phase space bins with the contribution of 
mixing taken into account is
\begin{equation}
\begin{split}
  M^+_{ij}{}' = & k_i K_{-j}+k_{-i}K_j \\
  +& 2\sqrt{k_ik_{-i}K_j K_{-j}}(c_iC_j + s_iS_j) \\
  +& 2K_j\sqrt{k_ik_{-i}}(y_Dc_i-x_Ds_i) \\
  +& 2K_{-j}\sqrt{k_ik_{-i}}(y_Dc_i+x_Ds_i) \\
  +& 2k_i\sqrt{K_j K_{-j}}(y_DC_j-x_DS_j) \\
  +& 2k_{-i}\sqrt{K_j K_{-j}}(y_DC_j+x_DS_j) \\
  +& O(x_D^2, y_D^2)
\end{split}
\label{m_mix_plus}
\end{equation}
for $\cconj=+1$, and 
\begin{equation}
\begin{split}
  M^-_{ij}{}' = & k_i K_{-j}+k_{-i}K_j \\
  -& 2\sqrt{k_ik_{-i}K_j K_{-j}}(c_iC_j + s_iS_j)\\
  +& O(x_D^2, y_D^2)
\end{split}
\label{m_mix_minus}
\end{equation}
for $\cconj=-1$. 

Important special cases for the two-body decays are: 
\begin{enumerate}
  \item Flavor-specific decay ({\it e. g.} $D^0\to K^-e^+\nu_e$): 
        $K_1=1$, $K_{-1}=0$,
  \item $D^0\to K^-\pi^+$ decay: $K_1=1$, $K_{-1}=r^2_{K\pi}$, 
        $C_1=\cos\delta_D$, $S_1=\sin\delta_D$. 
\end{enumerate}

The effect of mixing is linear in the case of $\cconj=+1$, 
which allows to measure the parameters $x_D$ and $y_D$ in the combined 
fit of the decays of both charge parities. The simplest strategy 
involves flavor-tagged \dkpp decays produced in both $\cconj=+1$ and 
$\cconj=-1$ states. Decay of the $\cconj=-1$ state yields the numbers of events 
in \dkpp phase space bins unaffected by mixing, while the $\cconj=+1$
state contains linear mixing contribution: 
\begin{equation}
  M^+_i{}'= K_{-i} + 2\sqrt{K_iK_{-i}}(y_DC_i-x_DS_i) + O(x_D^2, y_D^2). 
  \label{m_mix_tag}
\end{equation}
Since there are bins with $|C_i|\sim 1$ and bins with $|S_i|\sim 1$, 
the sensitivity to both $x_D$ and $y_D$ should be of the same order.
Note that the mixing term is twice larger with respect to that
in \dsdpi, \dkpp decays (Eq.~\ref{k_mix}). 

We have made a quantitative estimate of the sensitivity to mixing parameters 
using MC simulation. The results of this study for a data sample equivalent 
to the integrated luminosity of $10^{3}$ fb$^{-1}$ are shown in 
Table~\ref{tab:Mix_Error}. 
This corresponds to one standard year of data taking with a peak 
luminosity of $10^{35}$ cm$^{-2}$s$^{-1}$. 
Projects of $c\tau$-factories with such luminosity are being 
considered \cite{c_tau_nsk}.
The numbers of events are estimated using the double-tagged event 
efficiency obtained by CLEO in the reconstruction of $\psi(3770)$ 
decays~\cite{cleo_dkpp_corr} (see Table~\ref{tab:EvNum_Step} 
in Appendix~\ref{num_calc}).

\begin{table}
\begin{center}
\caption{Statistical sensitivity to the mixing and \cpconj 
violation parameters for the time-integrated Dalitz plot analysis using a
$10^{3}$ fb$^{-1}$ $e^+e^-\to \psi(4040)$ data sample. }
\label{tab:Mix_Error}
\begin{tabular}{|l|c|c|c|}\hline
Parameter & \multicolumn{3}{|c|}{Precision} \\ \cline{2-4}

&Coherent only&Incoherent only&Both\\ \hline
$x_D$ ($10^{-4}$)  			& $12.5$ & $18.4$ & $11.8$ \\
$y_D$ ($10^{-4}$)  			& $8.7$ & $12.9$ & $8.5$ \\
$r_{\cpconj}$ ($10^{-2}$)     		& $5.4$ & $5.2$ & $3.8$ \\
$\alpha_{\cpconj}$ (${}^{\circ}$) 	& $3.5$ & $3.5$ & $2.5$      \\ 
\hline
\end{tabular}
\end{center}
\end{table}

Our study involves only the flavor-tagged and double-tagged 
\dkpp decays in the coherent $\cconj=-1$ state 
(to extract $K_i$ and $C_i,S_i$, respectively), as well as the 
flavor-tagged \dkpp decays in the $\cconj=+1$ state and the incoherent 
flavor-tagged \dkpp decays for
the final extraction of mixing parameters.
The incoherent events can be obtained at a charm factory
from $\psi(4040) \to D^{\pm}D^{*\mp}$ decays. Table~\ref{tab:Mix_Error}
shows that adding the incoherent decays to the coherent sample does not 
improve the mixing parameters precision significantly.
It is explained by the fact that the precision is determined by 
the flavor-tagged \dkpp decays in $\cconj=-1$ state. 
However, the inclusion of incoherent decays improves the precision of 
\cpconj violation parameters.

A more complicated analysis can also include the 
double-tagged \dkpp decays in $\cconj=+1$ state which also include the 
mixing parameters at the first order.
Our estimation of the mixing parameters precision is based on 
the assumption that systematic errors can be essentially suppressed 
since both $\cconj=+1$ and $\cconj=-1$ decays have similar kinematics 
and may be detected simultaneously in the experiment.

Equations (\ref{m_mix_plus}) and (\ref{m_mix_minus})
imply the absence of direct \cpconj violation in $D$ decays. 
It is, however, possible to take its effect into account. This requires 
doubling the number of parameters $K_i$, $C_i$, $S_i$ (by treating 
decays of \dn and \dnbar separately for self-conjugate final states like 
\kspp, or by decoupling the numbers of events in \cpconj-conjugate 
modes for final states like $K^-\pi^+\pi^0$). 
The number of equations used to constrain the phase terms 
$C_i, S_i$ from the double-tagged $\dkpp$ decays will stay the same, 
but for a reasonably large number of bins the system of equations 
will remain solvable. Consequently, the method can be used to 
distinguish between direct \cpconj violation and \cpconj 
violation in mixing. 

\cpconj violation in mixing of neutral kaons can mimic \cpconj
violation in charm if the final state $K^0_S\pi^+\pi^-$ or other 
states including $K^0_S$ mesons are used in the analysis. \cpconj-violating 
terms in the amplitudes $A_{1,2}$ are of the order $\epsilon\lambda^2$, 
where $\epsilon\simeq 2.2\times 10^{-3}$ \cite{pdg} is the magnitude of 
\cpconj violation in kaon mixing, and $\lambda^2=\sin^2\theta_C\simeq 0.23^2$
gives the relative difference between the $D^0\to K^0_S\pi^+\pi^-$ 
and $D^0\to K^0_L\pi^+\pi^-$
amplitudes. If not accounted for, \cpconj violation in the kaon sector 
will introduce fake \cpconj violation in charm of the order 
$r_{CP}\sim \epsilon\lambda^2/(x_D^2+y_D^2)\sim 1\%$. However, 
once the amplitude of the $D^0\to K^0_L\pi^+\pi^-$ decay is measured
(this can be done at a charm factory exploiting kinematic reconstruction of
the missing $K^0_L$ at $D\overline{D}$ threshold), this effect can be 
corrected for in the charm mixing measurement.

\section{Extension to other $\mathbf{\dn}$ decay modes}

\label{other_modes}

The precision of the charm mixing parameters measurements can be 
improved by adding other three and four-body hadronic final states 
such as $K^-\pi^+\pi^0$ and $K^-\pi^+\pi^-\pi^+$ in addition to \dkpp. 
In these cases the phase terms $C_i$, $S_i$ can be defined in the same 
way as for \dkpp.

The advantage of using flavor-tagged Cabibbo-favored and doubly Cabibbo-suppressed 
$D^0\to K^{\mp}\pi^{\pm}\pi^0$
three-body decays for the mixing measurement is that the relative size of the 
interference term can be made much larger than in the case of \dkpp 
decays. For example, if we observe decays where one of the 
$D$ mesons is tagged by its semileptonic decay to be $D^0$, and the 
other one is reconstructed in the $K^-\pi^+\pi^0$ state (``wrong-sign'' tag), 
the first term in Eq.~\ref{m_mix_tag} is of the order 
$R_{K\pi\pi}\sim 0.06\times 0.06$, 
while the interference term is of the order 
$\sqrt{R_{K\pi\pi}}(x_D,y_D)\sim 0.06\times 0.01$. While the statistical sensitivity 
to the mixing parameters should be similar compared to the 
previous analysis, the systematic 
error can be reduced since the relative magnitude of the interference term that 
contains the mixing parameters is larger, and thus is less sensitive to background 
and efficiency uncertainties. 

Since the statistical sensitivity depends on the structure of the 
decay amplitude, we perform a MC study 
for the $D^0\to K^{\mp}\pi^{\pm}\pi^0$ final state. The description 
of CF and DCS $D^0\to K^{\mp}\pi^{\pm}\pi^0$ decay amplitudes is based on 
quasi two-body models obtained by CLEO~\cite{Kpipi0CFModel} and 
BaBar~\cite{Kpipi0DCSModel}.

\begin{figure}
\includegraphics[angle=90,width=0.45\textwidth]{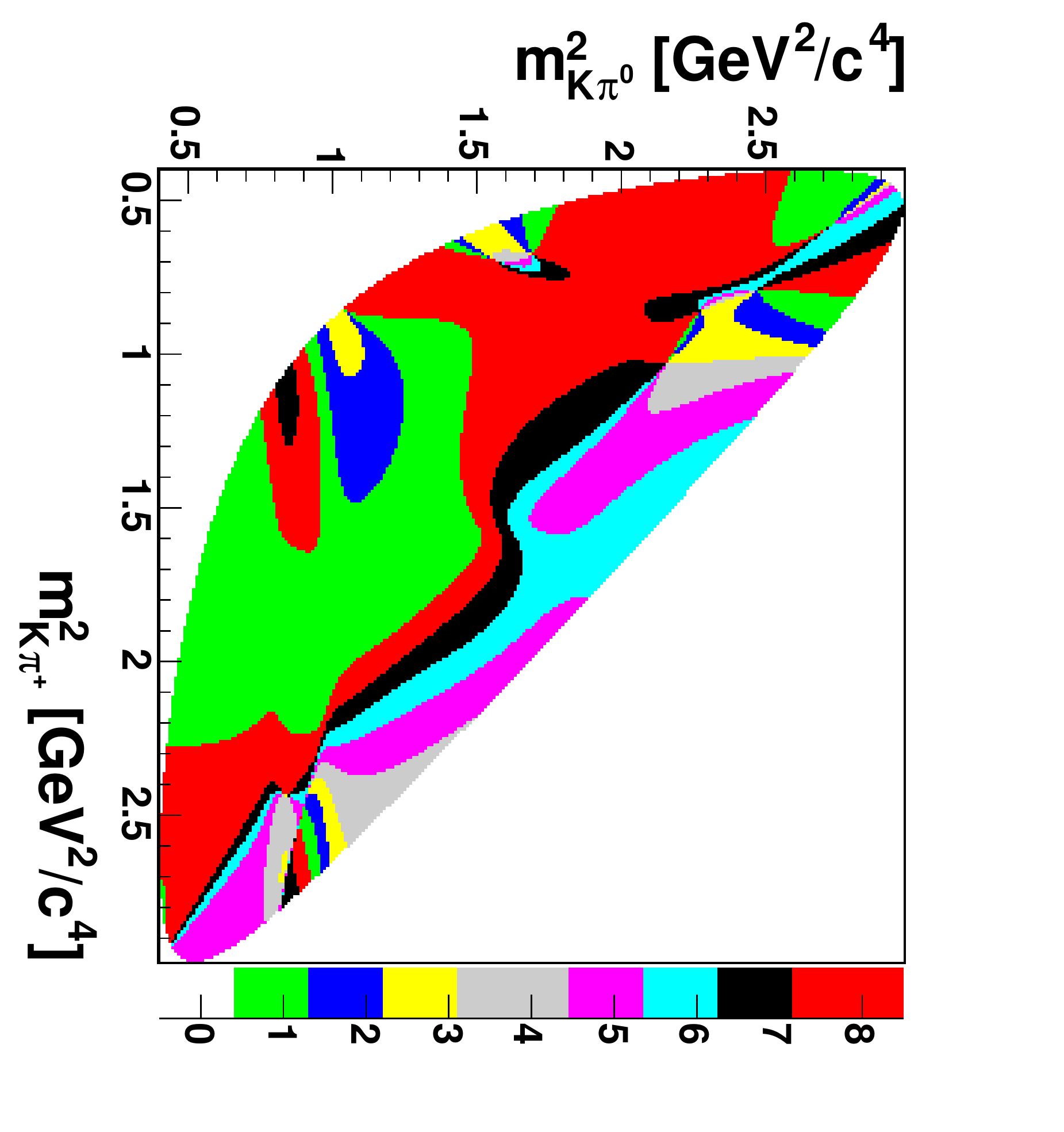}
\caption{Uniform phase binning obtained from the CF and DCS $D^0\to K^{\mp}\pi^{\pm}\pi^0$ decays amplitude models.}
\label{Kpipi0Binning}
\end{figure}

\begin{figure}
\vspace*{-0.05\textwidth}
\includegraphics[angle=90,width=0.45\textwidth]{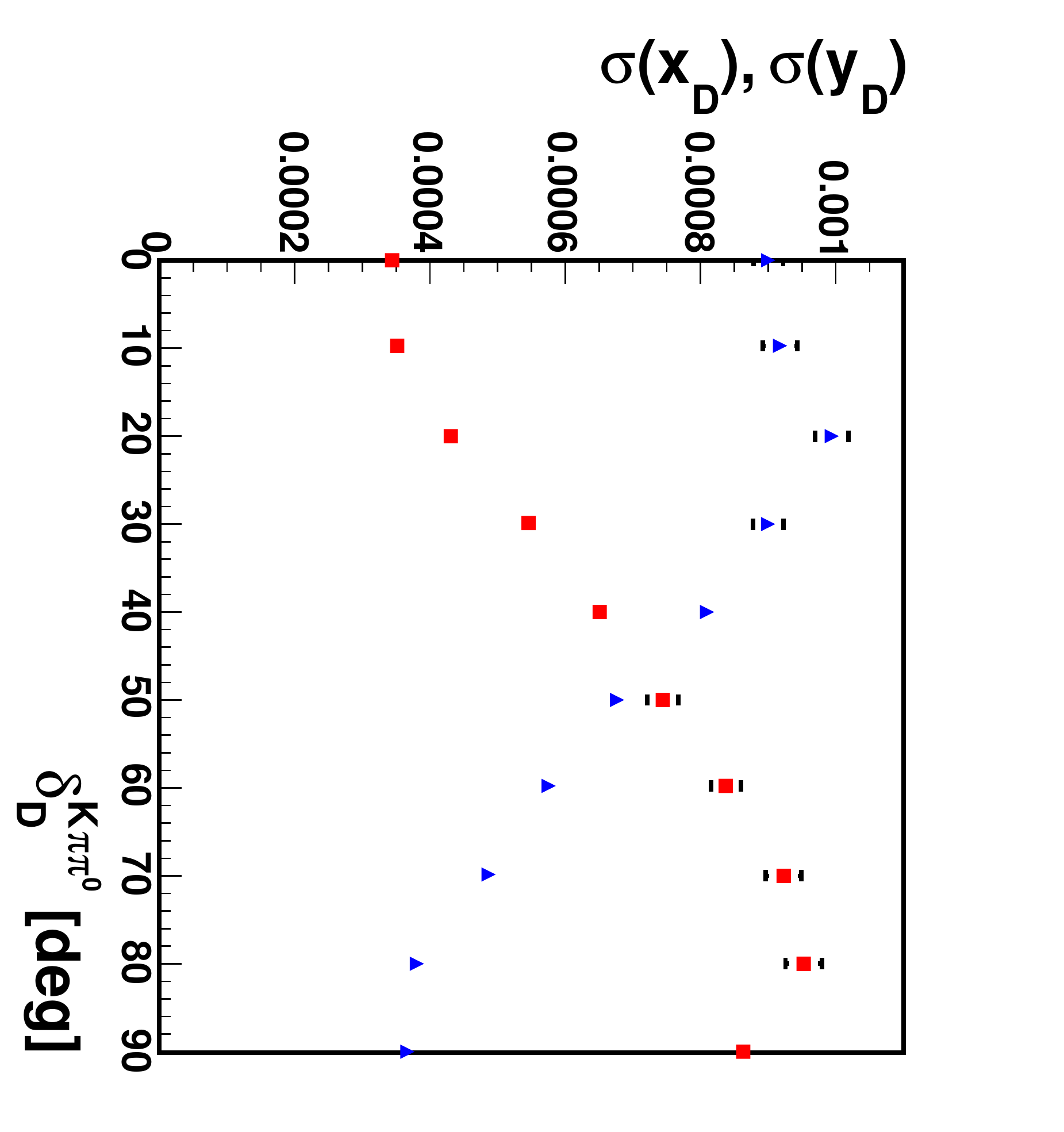}
\caption{Statistical precision for $x_D$ (triangles) and $y_D$ (squares) as functions of the average strong phase difference between CF and DCS $D^0\to K^{\mp}\pi^{\pm}\pi^0$ decay amplitudes.}
\label{Kpipi0results}
\end{figure}

As in the case of the $K_S^0\pi^+\pi^-$ decay, we use a uniform phase 
binning~\cite{modind2008}; Fig.~\ref{Kpipi0Binning} shows the binning 
obtained from the CF and DCS models used in our study. We use a sample 
corresponding 
to an integrated luminosity of $10^3$ fb$^{-1}$ (as for the $\dkpp$ decay).
The results of our 
calculation of the statistical uncertainties of the mixing parameters
as functions of the average strong 
phase difference between CF and DCS decay amplitudes $\delta_D^{K\pi\pi^0}$
are shown in Fig.~\ref{Kpipi0results}. 
Apparently, there is a symmetry: 
$\sigma_{x_D}\left (\delta_D^{K\pi\pi^0}\right ) = \sigma_{y_D}\left (\delta_D^{K\pi\pi^0}\pm\pi/2\right )$.
The strong phase difference obtained by CLEO equals $227^{+14}_{-17}{}^{\circ}$. 
Our estimates of the mixing parameters' precision for this 
value are shown in Table~\ref{tab:Mix_Error_Kpipi0}.

\begin{table}
\begin{center}
\caption{Statistical sensitivity to the mixing and \cpconj violation 
 parameters 
 using $D\to K\pi\pi^0$ decays. Results of toy MC simulation for 
 the integrated luminosity of $10^3$ fb$^{-1}$ and $\delta_D^{K\pi\pi^0}=227^{\circ}$.}
\label{tab:Mix_Error_Kpipi0}
\begin{tabular}{|l|c|c|c|}\hline
Parameter & \multicolumn{3}{|c|}{Precision} \\ \cline{2-4}

&Coherent only&Incoherent only&Both\\ \hline
$x_D$ ($10^{-4}$)  			& $6.2$ & $8.7$ & $6.0$ \\
$y_D$ ($10^{-4}$)  			& $6.1$ & $8.9$ & $6.1$ \\
$r_{\cpconj}$ ($10^{-2}$)     		& $2.6$ & $2.2$ & $1.7$ \\
$\alpha_{\cpconj}$ (${}^{\circ}$) 	& $2.4$ & $2.4$ & $1.7$  \\ 
\hline
\end{tabular}
\end{center}
\end{table}

\section{Conclusion}

We have investigated the impact of charm mixing on all stages of 
the binned model-independent measurement of the CKM phase $\gamma$ 
from \bdk, \dkpp decays. We show that ignoring the 
mixing at all stages of the analysis is safe at the current 
statistical accuracy: the bias on $\gamma$ with the currently measured 
mixing parameters is of order $0.2^{\circ}$. 

A model-independent approach to perform a time-dependent Dalitz plot analysis 
of the \dkpp decay to extract charm mixing parameters was considered. The 
approach is promising for high-statistics analyses at the LHCb experiment
and at SuperB factories.

We find that in the coherent production of the $\dn \overline{D}{}^{*0}$ 
system in 
$e^+e^-$ collisions, the effect of charm mixing can be enhanced if 
the $\overline{D}{}^{*0}$ is reconstructed in the $D^0\gamma$ decay, 
which requires the $\dn\dnbar$ system to be in a $\cconj=+1$ state.  
A model-independent method to measure charm mixing parameters from  
time-integrated analysis at a charm factory is proposed. Using a 
MC simulation we find that the 
sensitivity to both mixing parameters $x_D$ and $y_D$ is of order 
$10^{-3}$ for one year of data taking with luminosity 
$10^{35}$ cm$^{-2}$s$^{-1}$ at the peak of $\psi(4040)$ resonance. 
The method does not rely on absolute branching fraction measurements, 
and therefore does not contain uncertainties related to 
measurements of absolute efficiency or values of the branching fractions 
of the decays involved. 
Due to strong phase variations over the phase space of the 
\dkpp decay, nearly equal sensitivity to both $x_D$ and $y_D$ is 
obtained. 
The proposed method can also be used to measure the \cpconj violation
parameters of charm mixing. 

\section*{Acknowledgments}
We are grateful to David Asner, Alexey Garmash, Tim Gershon, Jonas Rademacker, 
and Guy Wilkinson for fruitful discussions and valuable comments.
We acknowledge support from 
the Ministry of Science and Education of the Russian Federation, 
the University of Warwick Research Development Fund, and the Science and Technology 
Facilities Council (United Kingdom). 


\appendix

\section{Formalism for coherent $\mathbf{\dn}$ decays with charm mixing}
\label{full_formalism}

We denote the amplitude of \dn (\dnbar) decay to the final state $f$ as 
\ad (\adbar). In the case of decay to three spin-zero particles, an 
amplitude \ad depends on two kinematic parameters (here we use two 
squared invariant masses $m_{12}^2$ and $m_{13}^2$ of the decay products). 
The density of the \dn decay Dalitz plot in the absence of charm mixing is
\begin{equation}
 \pd\left (m_{12}^2,m_{13}^2\right ) = \left |\ad\left (m_{12}^2,m_{13}^2\right )\right |^2.
\end{equation}
Charm mixing modifies this expression as
\begin{equation}
 \pd^{\prime}\left (m_{12}^2,m_{13}^2,t\right ) = 
   \left |\varkappa\left (t\right )\ad +
          i\frac{q}{p}\sigma\left (t\right )\adbar\right |^2, 
\end{equation}
where $t$ is the time between the production and the decay. 
Here we omit the dependence of 
$\ad$ and $\adbar$ on the Dalitz plot variables. 
The time-dependent terms are given by
\begin{equation}
\begin{split}
  \varkappa(t) + i\sigma(t) = \exp\frac{\Gamma t}{2}(-1+x_D-iy_D). 
\end{split}
\end{equation}

After integration over time we get
\begin{equation}
\begin{split}
\pd^{\prime}\left (m_{12}^2,m_{13}^2\right )=&a_0\pd + a_1r^{2}_{\cpconj}\pdbar +\\
               &r_{\cpconj}\sqrt{\pd\pdbar}\left (C^+a_2+S^+a_3\right )
\end{split}
\end{equation}
where
\begin{equation}
\begin{split}
 &r_{\cpconj}e^{i\alpha_{\cpconj}}=\frac{q}{p},\\
 &S^{\pm}=\sin{\left (\Delta\delta_D\pm\alpha_{\cpconj}\right )},\\
 &C^{\pm}=\cos{\left (\Delta\delta_D\pm\alpha_{\cpconj}\right )},
\end{split}
\end{equation}
$\Delta \delta_D=\left (\delta_D\left (m_{12}^2,m_{13}^2\right )-\bar \delta_{D}\left (m_{12}^2,m_{13}^2\right )\right )$ 
is a difference of strong phases for the decays $D^0\to f$ and $\bar D^0\to f$, and 
\begin{equation}
  \begin{split}
  a_0&=\frac{1}{2}\left (\frac{1}{1-y_D^2}+\frac{1}{1+x_D^2}\right )\\
     &=1+\frac{1}{2}\left (-x_D^2+y_D^2\right )+O\left ((x_D+y_D)^3\right ),\\
  a_1&=\frac{1}{2}\left (\frac{1}{1-y_D^2}-\frac{1}{1+x_D^2}\right )\\
     &=\frac{1}{2}\left (x_D^2+y_D^2\right )+O\left ((x_D+y_D)^3\right ),\\
  a_2&=\frac{y_D}{1-y_D^2}=y_D+O\left ((x_D+y_D)^3\right ),\\
  a_3&=\frac{x_D}{1+x_D^2}=x_D+O\left ((x_D+y_D)^3\right ).
 \end{split}
\end{equation}
The expressions for \dnbar decays can be obtained after the substitutions 
$p\leftrightarrow q$ and $\pd \leftrightarrow \pdbar$:
\begin{equation}
\begin{split}
\pdbar^{\prime}\left (m_{12}^2,m_{13}^2\right )=&a_0\pdbar + a_1r^{-2}_{\cpconj}\pd +\\
               &r^{-1}_{\cpconj}\sqrt{\pd\pdbar}\left (C^-a_2+S^-a_3\right )
\end{split}
\end{equation}

Now we consider the decay of the coherent $\dn\dnbar$ pair in a $\cconj=-1$ or $\cconj=+1$ state. 
Assuming the particle denoted with the index ``1'' decayed first, the Dalitz plot density 
accounting charm mixing effects is given by the expression

\vspace{3cm}

\begin{widetext}
\begin{equation}
 \begin{split}
 \pd_{corr}^{\cconj}&\left (\left (m_{12}^2\right )_1,\left (m_{13}^2\right )_1,\left (m_{12}^2\right )_2,\left (m_{13}^2\right )_2\right )=
  b^\cconj_0\left [\pd_1\pdbar_2+\pdbar_1\pd_2 + 2 \cconj \sqrt{\pd_1\pdbar_1\pd_2\pdbar_2}\left (C_1C_2+S_1S_2\right )\right ]\\
  +&b^\cconj_1\left [r^{-2}_{\cpconj}\pd_1\pd_2+r_{\cpconj}^2\pdbar_1\pdbar_2 + 2\cconj\sqrt{\pd_1\pdbar_1\pd_2\pdbar_2}\left (C^+_1C^+_2-S^+_1S^+_2\right )\right ]\\
  +&b^\cconj_2\left [\sqrt{\pd_2\pdbar_2}C^+_2\left (r_{\cpconj}\pdbar_1+r^{-1}_{\cpconj}\pd_1\right )+\cconj\sqrt{\pd_1\pdbar_1}C^+_1\left (
  r_{\cpconj}\pdbar_2+r^{-1}_{\cpconj}\pd_2\right )\right ]\\
  +&b^\cconj_3\left [\sqrt{\pd_2\pdbar_2}S^+_2\left (r_{\cpconj}\pdbar_1-r^{-1}_{\cpconj}\pd_1\right )+\cconj\sqrt{\pd_1\pdbar_1}S^+_1\left (
  r_{\cpconj}\pdbar_2-r^{-1}_{\cpconj}\pd_2\right )\right ],
 \end{split}
\end{equation}
where
\begin{equation}
 \begin{split}
  b_0^\cconj&=\frac{1}{2}\left [\frac{1+\cconj y_D^2}{\left (1-y_D^2\right )^2}+\frac{1-\cconj x_D^2}{\left (1+x_D^2\right )^2}\right ]\approx a_0+\frac{\cconj+1}{2}\left (-x_D^2+y_D^2\right ),\\
  b_1^\cconj&=\frac{1}{2}\left [\frac{1+\cconj y_D^2}{\left (1-y_D^2\right )^2}-\frac{1-\cconj x_D^2}{\left (1+x_D^2\right )^2}\right ]\approx \left (\cconj+2\right )a_1,\\
  b_2^\cconj&=\frac{\left (1+\cconj\right )y_D}{\left (1-y_D^2\right )^2}\approx\left (1+\cconj\right )a_2,\\
  b_3^\cconj&=\frac{\left (1+\cconj\right )x_D}{\left (1+x_D^2\right )^2}\approx\left (1+\cconj\right )a_3,
 \end{split}
\end{equation}
and $\cconj=\pm 1$ for the symmetric and antisymmetric states.
Note that in the case $\cconj=-1$ the interference term vanishes~\cite{xing}, 
while for $\cconj=+1$ it is doubled compared to the incoherent case.

The expression for the $\dkpp$ decay density in the 
$B^{\pm}\to DK^{\pm}$ process accounting for the charm mixing contribution is
\begin{equation}\label{eq:pBMixCP}
 \begin{split}
  P^{\prime}_{B^{\pm}}\left (m_{\pm}^2,m_{\mp}^2\right )&=
    a_0\left [\pd+r_B^2\pdbar+2\sqrt{\pd\pdbar}\left (x_BC+y_BS\right )\right ]\\
  &+a_1\left [r^{\pm 2}_{\cpconj}\pdbar+r^{\mp 2}_{\cpconj}r_B^2\pd + 
    2\sqrt{\pd\pdbar}\left (x_{B^{\pm}}^{\pm}C^{\pm}-
                            y_{B^{\pm}}^{\pm}S^{\pm}\right )\right ]\\
  &+a_2\left [x_{B^{\pm}}^{\pm}\left (r_{\cpconj}^{\pm 1}\pdbar+r^{\mp 1}_{\cpconj}\pd\right )+
    \sqrt{\pd\pdbar}\left (r_{\cpconj}^{\pm 1}+
                           r^{\mp 1}_{\cpconj}r_B^2\right )C^{\pm}\right ]\\
  &+a_3\left [y_{B^{\pm}}^{\pm}\left (r_{\cpconj}^{\pm 1}\pdbar-r^{\mp 1}_{\cpconj}\pd\right )+
    \sqrt{\pd\pdbar}\left (r_{\cpconj}^{\pm 1}-
                           r^{\mp 1}_{\cpconj}r_B^2\right )S^{\pm}\right ],
 \end{split}
\end{equation}
where $x_B^{\pm}=r_B\cos{\left (\delta_B\pm\gamma\pm\alpha_{\cpconj}\right )}$, 
$y_B^{\pm}=r_B\sin{\left (\delta_B\pm \gamma\pm\alpha_{\cpconj}\right )}$.
\end{widetext}

\section{Numerical calculation procedure}
\label{num_calc}

The numerical results presented in Sections~\ref{gamma_mixing}, \ref{mixing_tint} and
\ref{other_modes} are obtained using a procedure described below. 

In the model-independent approach, the phase space of the decay is 
divided into bins and we deal with numbers of events in each bin. 
The maximum likelihood method is used to obtain the parameters of 
interest from these numbers. The likelihood function is defined as
\begin{equation}\label{eq:lhFunc}
 -2\log{\mathcal L}=-2\sum_i{\log{P\left (X_i,\langle X_i \rangle\right )}},
\end{equation}
where $P\left (X,\langle X\rangle\right )$ is the Poisson probability 
function to observe $X$ events with an expected number of events 
$\langle X \rangle$.

\begin{table}
\begin{center}
\caption{Numbers of events used for mixing parameters measurement. 
The numbers for $K^-\pi^+\pi^0$ include both CF and DCS decays. }
\label{tab:EvNum_Step}
\begin{tabular}{|c|c|c|}\hline
Type & \multicolumn{2}{|c|}{Number of events ($10^6$)} \\ \cline{2-3}
& $K_S^0\pi^+\pi^-$ & $K^-\pi^+\pi^0$ \\ \hline
Incoherent flavor tagged& $6$ & $30$\\
Coherent $\cconj=-1$ flavor tagged& $2.1$ & $10.5$\\
Coherent $\cconj=+1$ flavor tagged& $1.4$ & $7$\\
Coherent $\cconj=-1$              & $0.6$ & $13$ \\
\hline
\end{tabular}
\end{center}
\end{table}

The observed numbers of events $X_i$ are obtained using the exact formulas 
with mixing (see Appendix~\ref{full_formalism}). When the systematic 
effect of charm mixing on the $\gamma$ measurement is studied, the values 
of $X_i$ are fixed (which corresponds to infinite statistics), while 
in the study of the statistical sensitivity of the mixing measurement in 
Sections \ref{mixing_tint} and \ref{other_modes}, we sample the $X_i$
values according to the Poisson distribution. 

The expected numbers of events $\langle X_i\rangle$ include the free 
parameters of interest ($x_B$ and $y_B$ in Section~\ref{gamma_mixing}
or mixing parameters in Sections \ref{mixing_tint} and \ref{other_modes}) 
and are expressed via the numbers of events in flavor-specific $D^0$ state $K_i$
and the phase terms $C_i$, $S_i$. The values of $K_i$, $C_i$ and $S_i$ are 
calculated from the \dkpp and $D^0\to K^{\mp}\pi^{\pm}\pi^0$ decay amplitudes
~\cite{belle_phi3,Kpipi0CFModel,Kpipi0DCSModel} as
\begin{equation}
  K_i=\int_{\mathcal D_i}{p_D dm_+^2dm_-^2}, 
\end{equation}
\begin{equation}
  C_i=\frac{\int_{\mathcal D_i}{\sqrt{p_D\bar p_D}
        \,C\, dm_+^2dm_-^2}}{
          \sqrt{\int_{\mathcal D_i}{p_D dm_+^2dm_-^2}
                \int_{\mathcal D_i}{\bar p_D dm_+^2dm_-^2}}}, 
\end{equation}
\begin{equation}
  S_i=\frac{\int_{\mathcal D_i}{\sqrt{p_D\bar p_D}
        \,S\, dm_+^2dm_-^2}}{
          \sqrt{\int_{\mathcal D_i}{p_D dm_+^2dm_-^2}
                \int_{\mathcal D_i}{\bar p_D dm_+^2dm_-^2}}}. 
\end{equation}
The free parameters are extracted using MINUIT to minimize the 
likelihood function~(\ref{eq:lhFunc}). 

Table \ref{tab:EvNum_Step} shows the numbers of
events that we use to obtain the results in Sections \ref{mixing_tint} and 
\ref{other_modes}. 
To estimate the numbers we use the detection efficiency determined 
by CLEO \cite{cleo_dkpp_corr} and the statistics corresponding to an integrated 
luminosity of $10^{3}$ fb$^{-1}$ at the $\psi(4040)$ resonance.

\end{document}